
\input harvmac
\noblackbox

\def\z{{\bf z}}
\def\b{{\bf b}}

\def\appA{A}

\lref\bvi{N. Berkovits and C. Vafa, \np {\bf 433} (1995) 123.}
\lref\bvii{N. Berkovits and C. Vafa, \np {\bf 533} (1998) 181.}
\lref\BL{S. Yau and E. Zaslow, {\it BPS states, string duality, and 
nodal curves on $K3$}, hep--th/9512121;\br
L. G\"ottsche,
A conjectural generating function for numbers of curves on
surfaces, alg-geom/9711012;\br
L.\ Bryan and N.\ Leung,
The enumerative geometry of K3 surfaces and modular forms, 
alg-geom/9711031.}
\lref\LSW{W. Lerche, S. Stieberger, N.P. Warner, {\sl Quartic Gauge Couplings
  from $K3$ Geometry}, hep-th/9811228;
{\sl Prepotentials from symmetric products}, hep-th/9901162.}
\lref\BS{L.\ Baulieu and S.\ Shatashvili,
{\sl Duality from topological symmetry,} hep-th/9811198.}
\lref\stieberg{S.\ Stieberger, \np {\bf 541} (1999) 109.}
\lref\sugra{A.\ Salam and E.\ Sezgin, \pl {\bf 154} (1985) 37; \br
M.\ Awada and P. K.\ Townsend, \pl {\bf 156} (1985) 51.}
\lref\BGMN{
M. Bianchi, E. Gava, F. Morales, K.S. Narain, 
\np {\bf 547} (1999) 96.}
\lref\costas{A. Gregori, C. Kounnas and P.M. Petropoulos, hep--th/9901117;
hep--th/9904151. }
\lref\MNVW{J.A. Minahan, D. Nemeschansky, C. Vafa, N. P. Warner, 
\np {\bf 527} (1998) 581, hep-th/9802168}
\lref\LS {W. Lerche and S. Stieberger, 
{\it Adv. Theor. Math. Phys.} {\bf 2} (1998) 1105.}
\lref\WL {W. Lerche, \np {\bf 308} (1988) 102.}
\lref\psw{K. Pilch, A. Schellekens, N.P. Warner, \pl {\bf 287} (1987) 362}
\lref\lsw {A. Schellekens, N.P. Warner, \pl {\bf 177} (1986) 317; \pl
{\bf 181} (1986) 339; \np {\bf 287} (1987) 87;
W. Lerche, B.E.W. Nilsson and A. N. Schellekens, \np {\bf 289}
(1987) 609; W. Lerche, B.E.W. Nilsson, A.N. Schellekens and N.P. Warner,
\np {\bf 299} (1988) 91; W. Lerche, A.N. Schellekens and N.P. Warner, 
\pr {\bf 177} (1989) 1.}
\lref\fs{K. F\"orger and S. Stieberger, {\it Nucl. Phys.} {\bf B 514} 
(1998) 135.}
\lref\ko{E. Kiritsis and N. Obers, {\it J. High Energy Phys.} 10 (1997) 4.}
\lref\PK{E. Kiritsis and B. Pioline, \np {\bf 508} (1997) 509.}
\lref \DKLII {L. Dixon, V. Kaplunovsky and J. Louis, \np {\bf 355} (1991) 
649.}
\lref\GSW{M. B. Green, J.H. Schwarz, E. Witten, {\sl Superstring
Theory},Cambridge U. Press}
\lref\MS{J. Maharana, J.H. Schwarz, \np {\bf 390} (1993) 3}
\lref\dWKLL{B. de Wit, V. Kaplunovsky, J. Louis, D. L\"ust, \np {\bf
451} (1995) 53.}
\lref\LLT{J. Lauer, D. L\"ust, S. Theisen, \np {\bf 309} (1988) 771}
\lref\BFKOV{C. Bachas, C. Fabre, E. Kiritsis, N.A. Obers, P. Vanhove, 
\np {\bf 509} (1998) 33.}
\lref\bk{C. Bachas, E. Kiritsis, Nucl. Phys. Proc. Suppl. 55 B (1997) 194.}
\lref\ba{C. Bachas, Talk at STRINGS'97 (Amsterdam, June 16-21) and 
HEP-97 (Jerusalem, August 19-26), hep-th/9710102}
\lref\DMVV{R. Dijkgraaf, G. Moore, E. Verlinde, H. Verlinde, \cmp {\bf
185}(1997) 197}
\lref\Serre{J. P. Serre, {\it Cours d'Arithmetique}, PUF Paris 1970}
\lref\wie{E. Witten, \np {\bf  443} (1995) 85.}
\lref\pw{J. Polchinski, E. Witten, \np {\bf 460}(1996) 525.}
\lref\senu{A. Sen, Nucl. Phys. Proc. Suppl. 58 (1997) 5.}
\lref\sen{A. Sen, \np {\bf 475} (1996) 562}
\lref\ab{I. Antoniadis, C. Bachas, C. Fabre, H. Partouche, T. R. Taylor,
\np {\bf 489} (1997) 160}
\lref\w{E. Witten, {\sl Toroidal compactification without vector
structure}, JHEP {\bf 02} (1998) 006.}
\lref\lowe{D. A. Lowe, \pl {\bf 403} (1997) 243;
P. Horava, \np {\bf 505} (1997) 84;
S. J. Rey, \np  {\bf 502} (1997) 170; 
D. Kabat, S. J. Rey, \np {\bf 508} (1997) 535.}
\lref\gn{E. Gava, J. F. Morales, K. S. Narain, G. Thompson, 
\np {\bf 528} (1998) 95.}
\lref\dvv{R. Dijkgraaf, E. Verlinde, H. Verlinde, \np {\bf 500} (1997) 43.}
\lref\vafa{C. Vafa, \np {\bf 469} (1996) 403.}
\lref\bi{M. Bianchi, \np {\bf 528} (1998) 73.}
\lref\ABPSS{C. Angelantonj, M. Bianchi, G. Pradisi, A. Sagnotti,
Y. S. Stanev, \pl {\bf 385} (1996) 96}
\lref\bisa{M. Bianchi, G. Pradisi, A. Sagnotti, \np {\bf 376}(1992) 365}
\lref\BF{C. Bachas, C. Fabre, \np {\bf 476} (1996) 418.}
\lref\BP{C. Bachas, M. Porrati, \pl {\bf 296} (1992) 77.}
\lref\ts{A. A. Tseytlin, \np {\bf 467} (1996) 383; 
\pl {\bf 367} (1996) 84}
\lref\g{P. Ginsparg, Phys. Rev {\bf D 35} (1987) 648}
\lref\nsw{K. S. Narain, M. H. Sarmadi, E. Witten, \np {\bf 279} (1987) 369}
\lref\Terras{A. Terras, {\sl Harmonic Analysis on Symmetric Spaces and
Applications} (1985) Springer-Verlag}
\lref\apt{M. Serone, \pl {\bf 395} (1997) 42; 
I. Antoniadis, H. Partouche, T. R. Taylor, \np {\bf 489}
(1997) 160; J. F. Morales, M. Serone, \np {\bf 501} (1997) 427.} 
\lref\anton{I. Antondiadis, E. Gava, K. S. Narain, \np {\bf 383}
(1992) 109; 
I. Antondiadis, E. Gava, K. S. Narain, T. R. Taylor,  \np {\bf 407}
(1993)  706}
\lref\agnt{I. Antondiadis, E. Gava, K. S. Narain, T. R. Taylor,
 \np {\bf 455} (1995) 109}
\lref\antlect{I. Antoniadis, H. Partouche, T. R. Taylor, 
Nucl. Phys. Proc. Suppl. 61 A (1998)
58-71; Nucl. Phys. Proc. Suppl. 67 (1998) 3-16}
\lref\MM{M. Marino, G.  Moore, \np {\bf 543} (1999) 592.}

\lref\van{N. Vanegas--Arbel\'aez, 
{\sl Regularization of automorphic functions of manifolds with special 
K\"ahler geometry}, hep--th/9906028.}
\lref\erdelyi{W. Magnus, F. Oberhettinger and F.G. Tricomi,
{\sl Higher Transcendental Functions}, volume 1 and 2, ed. A. Erd\'elyi, 
McGraw--Hill 1953.} 
\lref\KK{E. Kiritsis and  C. Kounnas, 
\np {\bf 442} (1995) 472}

\def\cmp {{\it Comm. Math. Phys.\ }}
\def\np {{\it Nucl. Phys.\ } {\bf B\ }}
\def\pl {{\it Phys. Lett.\ } {\bf B\ }}
\def\pr {{\it Phys. Rept.\ } }
\def\h {{1\over 2}}

\def\La {\Lambda}
\def\ov {\overline}
\def\o {\over}
\def\Li {{\cal L}i}
\def\Uc {{\ov U}}
\def\cP {{\cal P}}

\def\br{\hfill\break}
\def\IZ{ {\bf Z}}

\def\Ac {{\cal A}}
\def\Bc {{\cal B}}

\def\Fc {{\cal F}}
\def\Gc {{\cal G}}

\def\Kc {{\cal K}}

\def\Mc {{\cal M}}
\def\Nc {{\cal N}}
\def\Oc {{\cal O}}

\def\Tc {{\cal T}}
\def\Uc {{\cal U}}

\def\e{\epsilon}
\def\Bc {\cal B}
\def\tr {{\rm Tr}}
\def\det {{\rm det}}

\def\lf {\left}
\def\ri {\right}
\def\ra {\rightarrow}

\def\p {\partial}
\def\la{\langle}
\def\ra{\rangle}

\def\eps{\epsilon}
\def\tr{{\rm tr}}
\def\hf{{1\o 2}}


\Title{\vbox{\rightline{\tt hep-th/9901020} \rightline{CERN--TH/98--412}
\rightline{CPHT-S707.1298}}}
{\vbox{\centerline{
Higher Derivative Couplings and Heterotic--Type I} 
\centerline{Duality in Eight Dimensions}}}

\centerline{K. F\"orger$^1$ and S. Stieberger$^2$}

\bigskip\centerline{\it $^1$ Centre de Physique Th\'eorique}
\centerline{\it \'Ecole Polytechnique}
\centerline{\it F--91128 Palaiseau Cedex, FRANCE}
\bigskip\centerline{\it $^2$ CERN, Theory Division}
\centerline{\it CH--1211 Geneva 23, SWITZERLAND}
\bigskip

\centerline{\bf Abstract}
We calculate $F^4$ and $R^4T^{4g-4}$
couplings in $d=8$ heterotic and type~I string
vacua (with gauge and graviphoton field strengths $F,T$, 
and Riemann curvature $R$). 
The holomorphic piece $F_g$ of the heterotic one--loop coupling
$R^4T^{4g-4}$ is given by a polylogarithm of index $5-4g$
and encodes the counting of genus~$g$ curves with $g$ nodes 
on the $K3$ of the dual $F$--theory side.
We present closed expressions for world--sheet
$\tau$--integrals with an arbitrary number of lattice vector 
insertions.
Furthermore we verify that the corresponding heterotic one--loop 
couplings  sum up 
perturbative open string and non-perturbative D--string contributions
on the type I side.  
Finally we discuss a type I one--loop  correction to  the $R^2$ term.

\Date{12/98} 

\newsec{Introduction}

Heterotic $SO(32)$/type \ I duality relates 
two string theories in ten dimensions \wie\pw.
A field redefinition transforms the low energy effective action of the
heterotic string into the one of the type I string. This duality
map is valid for second derivative terms as well as 
for higher derivative terms containing four or eight derivatives
that can be arranged into superinvariants \ts.
The low energy effective actions 
of the heterotic and type I string theory
are related by the following field redefinitions in $d=10$ \pw\ts:
\eqn\dualmap{
G_{\mu\nu}^I= \lambda^I G_{\mu\nu}^{\rm het}\ \ \ \ 
\lambda^I={1\o\lambda^{\rm het}}\ \ \ \ 
B_{\mu\nu}^{R,I}=B_{\mu\nu}^{NS,{\rm het}}
\ \ \ \ A^{a,I}_{\mu}=A^{a,{\rm het}}_{\mu} }
where $a=1,\ldots,16$ labels
$SO(32)$ gauge indices, $G_{\mu\nu}$ is the metric and
$B_{\mu\nu}^{NS/R}$ the antisymmetric tensor in the $NS$ or $R$ sector
and  $\lambda^I=e^{\phi^{(10)}}$ is the ten dimensional type I coupling.
Compactifying to lower dimensions
modifies these relations due to the dependence on the
volume of the compactification manifold \ABPSS
\eqn\dil{
\phi_{\rm het}^d={6-d\o 4}\phi_I^d-{d-2 \o 16}\ln\det G_I}
where $G_I$ is the internal metric in the type I string frame.
This relation shows that the critical dimension $d=6$
separates  a strong--weak coupling duality 
from a weak--weak coupling duality.

Toroidal compactification down to eight dimensions
gives rise to $4$ Abelian gauge fields, 
corresponding to components of $G_{\mu I}$ and $B_{\mu
I}$, where $\mu$ is the space-time index and $I$ labels the
compact directions.
The four massless scalar fields from $G_{IJ}$ and $B_{IJ}$
combine into two complex
scalars $T,U$ and parametrize the moduli space
${O(2,2)\o O(2)\times O(2)}$ of the torus $T^2$
(if no Wilson lines are switched on).
In the type I theory the $U(1)^2\times U(1)^2$ gauge group is
reduced to a diagonal $U(1)^2$, due to the twist operator 
$\Omega$ \bisa. Similarly only two of the four massless scalar 
fields survive the projection.

Besides a comparison of the BPS spectrum of the heterotic and
type I string non--trivial checks can be done for the BPS--saturated 
${F}^4$ and ${R}^4$ couplings; see e.g. \bk\BFKOV\ko\BGMN\  
and \LS\LSW\ for gauge couplings in the context of  
heterotic/F-theory duality.
The special feature of these couplings is that --on the heterotic side-- 
they are protected from higher than one--loop corrections in the
effective action because of supersymmetry. Thus they can be exactly
calculated.
These threshold corrections translate to perturbative open string 
amplitudes and non-perturbative BPS $D1$-instanton corrections on  
the type\ I side.
Therefore by studying various examples of couplings 
in any dimension one gains insight into the rules of computing 
non-perturbative $D$-brane contributions.
It is the aim of this article to pursue these ideas further. 
The subject are eight-dimensional\foot{Other 
interesting aspects of $D=8$ theories have been recently discussed in \BS.}
heterotic and type I string vacua coming from toroidal compactification
of ten dimensions. We calculate various couplings of higher space--time 
derivatives and higher order in $\alpha'$ 
for both heterotic (section 2) and type I string vacua (section 3) 
and compare them: 
In section 2.1 we review the heterotic $R^4$ couplings, since they capture
a part of the structure of the $F^4$ couplings which originate in the 
torus compactification to eight dimensions. These couplings are calculated
in section 2.2. and 2.3. The latter section contains also 
closed expressions for world--sheet
$\tau$--integrals with an arbitrary number of lattice vector 
insertions. In section 2.4. we calculate the higher order corrections
$R^4T^{4g-4}$, which have a nice interpretation on the dual $F$--theory side.
The type I calculation of these couplings are presented in sections 3.1-3.3.
In section 3.4. we show that, a part of the heterotic $R^4T^{4g-4}$ 
couplings has an intriguing physical interpretation in term of the dual 
type I string picture as $D$--instanton sum.
Finally in section 3.5. we find a one--loop $R^2$ corrections in 
eight--dimensional type I string theory and give its explanation
on the heterotic side.

\newsec{Heterotic string in eight dimensions}

In this section we consider $T^2$ compactifications 
of the heterotic string.
The heterotic dilaton in eight dimensions is a real scalar in the 
supergravity multiplet \sugra. Therefore, it is believed that 
there exist no higher than one--loop
perturbative corrections, since by supersymmetry they would also affect 
the Einstein term. 
At string one--loop,  higher derivative couplings such as
${F}^4, ({F}^2)^2, {R}^4$ and $({R}^2)^2$ and 
their respective anomaly--cancelling $CP$--odd parts receive corrections
in the effective string action. These constant contributions are highly fixed
by ten--dimensional anomaly cancellation arguments.
In Green-Schwarz formalism they are calculated by (almost) holomorphic 
one loop string amplitudes, whose minimal number of external legs is fixed by 
saturating fermionic zero modes.
The result for the amplitudes  is summarized by
the worldsheet $\tau$ integral over a weight zero
almost holomorphic function which is related 
to the elliptic genus \lsw\WL.
After compactification on $T^2$ these corrections
become moduli dependent functions. It is believed that there are no
space--time instanton effects, since the only supersymmetric soliton,
the NS five--brane, cannot be wrapped around the torus. 
On the other hand, the non--anomaly related coupling to
$J_0=t_8t_8{R}^4-{1\o 8}\epsilon_{10}\epsilon_{10}{R}^4$ 
may receive higher perturbative as well as non--perturbative corrections.

\subsec{Ten dimensional origin}
Let us briefly review the four graviton amplitude.
The graviton vertex operator 
can be read off from the background field expansion of
the heterotic sigma model on the worldsheet torus \WL\psw \ and reads
in the Green-Schwarz formalism   
\eqn\grvert{
V_{\rm gr}=-{1\o 3}R_{\mu\rho\nu\sigma}\ :\bar\p X^\mu X^\rho\Big(\p X^\nu 
X^\sigma+{3\o 8}
\gamma_{ab}^{\nu\sigma} S^a S^b\Big):\ }
where $X^\mu$ are bosonic fields with $\mu=1\ldots 8$ 
and $S^a$ with $a=1\ldots 8$ their supersymmetric partners. 
The four-point graviton correlation function is
\foot{Using the vertex operator 
$V_{\rm gr}=\e_{\mu\nu}:\bar\p X^\mu(\p X^\nu-{\gamma^{\rho\nu}_{ab}\o 
4}k_\rho S^a S^b) e^{i k\cdot X}:$ the $\Oc(k^4)$ piece of the 
four point graviton amplitude produces the one-loop $\Tr R^2$  coupling 
with bosonic correlator
$\int \prod_{i=1}^4 d^2z_i \la \prod_{i=1}^4 \bar \p X^{\mu_i}\ra$, 
which gives no contribution
after $z$-integration \lsw.}
\eqn\ptgr{
\langle\prod_{i=1}^4 V^i_{\rm gr}\rangle={1\o 2^4}
t^{\nu_1\sigma_1\nu_2\sigma_2\nu_3\sigma_3\nu_4\sigma_4} 
 \prod_{i=1}^4R_{\mu_i\rho_i\nu_i\sigma_i}
 \langle\prod_{i=1}^4 :\bar\p X_i^{\mu_i}X_i^{\rho_i}:\rangle\ .}
The tensor 
$t_8\equiv t^{ijklmnpq}={\gamma_{ab}^{ij}\o 4}{\gamma_{cd}^{kl}\o 4}
{\gamma_{ef}^{mn}\o 4}{\gamma^{pq}_{gh}\o 4}\e^{abcdefgh}$ 
is defined e.g. as in \GSW. In the path integral  formalism it 
arises after integration over fermionic zero modes.

Depending on different ways of contracting bosonic fields 
the worldsheet integration results in two different couplings \lsw.
Contractions which are associated to 
permutation cycles $(1)(234)$ give rise to the $\Tr\ R^4$ coupling 
${\bar G}_4$, 
whereas contractions corresponding to permutation cycles $(12)(34)$ 
result in the $(\Tr R^2)^2$ coupling ${\hat {\bar G}}_2^2$.
The Eisenstein functions $\hat {G}_{2k}$  of weight $2k$ 
can be interpreted as `gravitational' charge insertions.\foot{The 
relation between
different normalized Eisenstein functions is
${G}_{2k}=-{(2\pi i)^{2k}\o (2k)!}B_{2k} E_{2k}=2\zeta(2k)E_{2k}$, 
where $B_{2k}$ are the Bernoulli numbers.}  

The pure gravitational part of the effective action  
is \lsw
\eqn\effact
{S_{\rm eff}^{\rm 1-loop}=-\Nc \int {d^2\tau\o
\tau_2^2} \tilde{\Ac}(\bar q,R)|_{8-form}\ ,}
where  $\Nc={V^{(10)}\o 2^{10}\pi^6}$  is the normalization factor
including the uncompactified volume 
and with the elliptic genus
\eqn\ellg{
\Ac(\bar q,0,R)=\exp\Big[\sum_{k=1}^\infty {1\o 4 k} {1\o (2\pi
i)^{2k}}
\Tr\Big({iR\o 2\pi}\Big)^{2k} \bar{G}_{2k}(\bar\tau)\Big] \ {\bar{E}_4^2\o
\bar{\eta}^{24}}\ ,}
where the last term is the $E_8\times E_8$ 
light--cone partition function $\Ac(\bar q)={1\o \bar{\eta}^{24}}
\sum_{P\in \Gamma_{16,0}}q^{P^2/2}$,
which is a modular form of weight $-4$. 
Since the effective action has to be modular invariant
the elliptic genus is replaced by its
regularized version $\tilde \Ac(q,R)$ depending on the
non-holomorphic
function ${\hat G_2}$ instead of $G_2$.

The action \effact\ has also to be supplemented
with the corresponding GS--counterterm which will provide us the $CP$--odd
part of our amplitudes and gives the eight derivatives gravitational
superinvariants $I_3, I_4$ \ts.

\subsec{$F^4$ heterotic couplings} 
For toroidal compactification the amplitudes
include a lattice sum, which arises from integration over
zero modes of compactified bosons. 
In addition to the gauge group $E_8\times E_8$ or $SO(32)$ 
(without additional Wilson lines) there are extra $U(1)$ gauge groups.
Here we will mainly concentrate on these $U(1)$ gauge couplings and
consider the case without and with Wilson lines separately.
The internal gauge boson vertex operator
can be read off from the $\sigma$ model action.
Writing the internal metric as
\eqn\metric{
G_{IJ}={T_2\o U_2}\left(\matrix{1&U_1\cr U_1& |U|^2}\right)
=\e_{(I} \bar\e_{J)}\ ,}
with $\e_I=\sqrt{T_2\o U_2}(1,U)$ and
$\bar\e_J=\sqrt{T_2\o U_2} (1,\bar U)$ 
the vertex operator for $G_{\mu I}$ gauge bosons 
in a background gauge 
$\e_{\mu}e^{i k\cdot X}=-\hf F_{\mu\rho} X^{\rho}$ with 
$F_{\mu\nu}=\rm const.$ in Green-Schwarz formalism is:
\eqn\vgauge{
V_{\rm gauge}={i\pi\o \tau_2} F_{\mu\nu}^A\ \bar Q_R^A
:\Big(\p X^\mu X^\nu-{1\o 4}
 S^a \gamma_{ab}^{\mu\nu} S^b\Big):}
where $A=1,2$ labels different $U(1)$ charges
\eqn\qr{\eqalign{ 
Q_R^1\equiv\bar Q_R&=\eps_I\bar\p X^I=\sqrt{T_2\o U_2} (1,U) A 
\left(\matrix{\tau\cr 1}\right)\ ,\cr
Q_R^2\equiv Q_R&=\bar\eps_I\bar\p X^I=\sqrt{T_2\o U_2} (1,\bar U) A 
\left(\matrix{\tau\cr 1}\right)\ ,}}
where $A=\Big({n_1\atop n_2} {-l_1\atop -l_2}\Big)$ 
is a $GL(2,{\bf Z})$ matrix. 
The complex structure and the K\"ahler
structure of the torus are defined in terms of the metric
and NS-NS antisymmetric tensor as $U=U_1+i U_2=(G_{12}+i\sqrt{\det 
G})/G_{11}$
and $T=T_1+iT_2=2 (b+i\sqrt{\det G})$, respectively.

Inspection of the four point amplitude 
$\la \prod_{i=1}^4V^{A_i}_{\rm gauge}\ra$
shows that replacing $R_{\mu\rho\nu\sigma}\bar \p X^\mu\ X^\rho$  
by $F_{\mu\nu,A} Q^A$ in \ptgr \ turns the four graviton amplitude into the
four gauge boson amplitude with bosonic correlator
$\la \prod_{i=1}^4 Q^A_{R,i} \ra_{\rm int}$.
This correlation function now includes a sum over
winding modes $\bf n$ and $\bf l$ of  the $\Gamma_{(2,2)}$ lattice.

The effective action 
depending on these couplings can be expressed as:
\eqn\hamp{\eqalign{
S^{\rm het}_{\rm 1-loop}& ={V^{(8)} T_2\o 2^{8} \pi^4}
\int {d^2\tau\o\tau_2^2} 
\sum_{{\bf l},{\bf n}\in {\bf Z}}e^{-{2\pi\o \tau_2}
(n^I\tau-l^I)(G+B)_{IJ}(n^J{\bar\tau}- l^J)}\cr
&\times \int DS^a_0 \ e^{{-\pi/\tau_2}
F_{\mu\nu,A}\ Q_R^A R_0^{\mu\nu}} \bar\Ac(\bar q) \ ,}}
where $R_0^{\mu\nu}={1\o 4}S_0^a \gamma_{ab}^{\mu\nu} S_0^b$.\foot{
Supersymmetry relates even to  odd spin structures. Since periodic
Green-Schwarz fields $S^a$ are mapped to periodic NSR field $\psi^i$, 
$CP$-odd correlation functions of NSR currents are equivalent to
$CP$-even correlation functions of Green-Schwarz currents due to
a Riemann identity {\WL}.}  

Berezin integration thus produces the following terms in the
effective action
\eqn\ihet{
{S}^{\rm het}_{\rm 1-loop}={V^{(8)}T_2\o 2^8 \pi^4}
\Delta_{F_1^{4-q} F_2^q}t_8 F_1^{4-q} F_2^q \ ,}
for $q=0,\ldots,4$ with coupling
\eqn\deltaa{\eqalign{
\Delta_{F_1^{4-q} F_2^q}&= 
{\partial^{4-q}\o \partial {\La_1}^{4-q}}{\partial^{q}\o\partial {\La_2}^{q}}
\times \cr
&\times \int {d^2\tau\o\tau_2^{2}} \sum_{A\in GL(2,{\bf Z})}
e^{2\pi i T \det A-{\pi T_2\o \tau_2 U_2}|(1,U) 
A\left(\tau\atop -1\right)|^2}
e^{- \La_A {\pi\o\tau_2} Q_R^A}\bar{\cal A}(\bar q)
\Big|_{\La=0} \  .}}
The gravitational charges $G_{2k}$ 
are now replaced by $\prod_{i=1}^4 {\pi\o \tau_2}Q^{A_{R,i}}$ 
which transforms under $SL(2,{\bf Z})_{\tau}$ as a modular function 
of weight $4$.
 
Poisson resummation turns the sum over winding modes ${\bf l}, {\bf
n}$ into
a sum over momenta ${\bf m}$ and windings ${\bf n}$. 
The partition function  
now contains Narain momenta insertions 
\eqn\narimp{\eqalign{
p_R&={1\o\sqrt{2 T_2 U_2}}[m_1+m_2\bar U+T(n_1
+n_2\bar U)]\cr
p_L&={1\o\sqrt{2 T_2 U_2}}[m_1+m_2\bar U+\bar T(n_1
+n_2\bar U)]\ .}}
The different couplings $\Delta_{F_1^{4-q} F_2^q}$
include charge insertions $(\bar Q_R)^{4-q} (Q_R)^q$.
From \deltaa \ the $\Tr F_1^{4-q} F_2^q$ coupling reads: 
\eqn\intFV{\eqalign{
\Delta_{F_1^{4-q} F_2^q}&=
\pi^4 \int {d^2 \tau \o \tau_2^6} 
\sum_{n_1,n_2 \atop  l_1,l_2}
(\bar Q_R)^{4-q} (Q_R)^{q}e^{-2\pi i \ov T \det A}  e^{{-{\pi T_2} 
\o {\tau_2
U_2}}\lf|n_1\tau+n_2U\tau -Ul_1+l_2 \ri|^2}
\ {\bar E_4^2\o \ov \eta^{24}}\ .\cr}}
This $\tau$-integral has been solved in \LS\ by the method
of orbit decomposition \DKLII\ and techniques developed in \fs.
It has been  shown in \LS \ that these four point functions \intFV\ 
are related
to a holomorphic prepotential $\Gc(T,U)$ which transform as
a modular function of weight $(-4,-4)$ under
$SL(2,{\bf Z})_T\times SL(2,{\bf Z})_U$ duality. 
This is similar to vector multiplet sector of N=2 supersymmetric 
string vacua in four dimensions.
They arrange as fourth order covariant 
derivatives\foot{The covariant derivative 
$D_w$ maps modular forms $\Phi_{w,\bar w}(U)$ of weight 
$(w,\bar w)$ to forms of
weight $(w+2,\bar w)$ i.e. 
$\Phi_{w+2,\bar w}(U)=D_w\Phi_{w,\bar w}(U)
 ={i\o\pi}(\p_U+{w\o (U-\bar U) })\Phi_{w,\bar w}(U)$.
We use the notation $D^k\Phi_{w,\bar w}=D_{w+2 (k-1)}
D_{w+2(k-2)}\ldots D_w \Phi_{w,\bar w}$.}
\eqn\RES{
\Delta_{F_1^{4-q} F_2^q}= 16\pi i\Big({T_2\o U_2}\Big)^{2-q}
D_T^{4-q} D_U^q\Gc-16\pi i\Big({U_2\o T_2}\Big)^{2-q} {\bar D}_{\ov T}^q
{\bar D}_{\ov U}^{4-q}\bar{\Gc}}
of the prepotential $\Gc$ \LS
\eqn\prep{
{\cal G}(T,U)=-{ic(0)\zeta(5)\o 64\pi^5}-{1\o 120}U^5
-{i\o (2\pi)^5}\sum_{(k,l)>0} c(kl)\ \Li_5 \big[{q_T}^k {q_U}^l\big]+
Q(T,U)\ ,}
with some quartic unconstrained polynomial $Q(T,U)$. 
The holomorphic covariant coupling $\p_T^5 \Gc$ is obtained by
a five point amplitude $\la V_T \prod_i^4 V^{(i)}_{F_T}\ra$ \LS.
This is the analogue to the Yukawa couplings $\p_T^3 f$ in the $N=2$
supersymmetric case in $d=4$. 

\subsec{Generalized worldsheet $\tau$-integrals}

Rather than restricting to the case \intFV, it is the purpose of this
section to find general expressions for world--sheet $\tau$--integrals
with an arbitrary number of lattice vector $\ov p_R, p_R, \ov p_L$ and 
$p_L$ insertions.
Such integrals appear in $N$--point string amplitudes of toroidal
string compactifications. See also \fs\ for further discussions
and \fs\LS\MM\agnt\ for examples.
\eqn\setup{\eqalign{
\Delta_{q_1q_2q_3q_4}&:=
{\p^N \o \p\Lambda_1^{q_1} \p\Lambda_2^{q_2} \p\Lambda_3^{q_3} 
\p\Lambda_4^{q_4}} 
\int {d^2\tau \o \tau_2}\ \tau_2^r  {1\o \tau_2^s} 
\sum_{(p_L,p_R)}\ e^{\pi i\tau|p_L|^2} e^{-\pi i\ov\tau|p_R|^2}\cr
&\times \lf. 
e^{\Lambda_1\ov p_R+ \Lambda_2p_R +\Lambda_3\ov p_L+\Lambda_4 p_L}\ 
e^{-{1\o 2\pi\tau_2}(\Lambda_1\Lambda_2+\Lambda_2\Lambda_3+\Lambda_3\Lambda_4+
\Lambda_4\Lambda_1)}\ \ov f_{k}(\ov \tau)\ri|_{\Lambda_i=0}
\ ,\cr}} 
with the modular function $f_k$ of weight $k$,
the integers $r,s\geq 0$,\ 
$q_1,q_2,q_3,q_4\geq 0$ and $N=q_1+q_2+q_3+q_4$.
We exclude the trivial case $(q_1,q_2,q_3,q_4)=(0,0,0,0)$.
World--sheet modular invariance of the integrand requires:
\eqn\conds{\eqalign{
k&=-q_1-q_2+q_3+q_4\cr
r-s-k&=q_1+q_2\ .\cr}}
In particular, the expression 
$(T- \ov T)^{-m} (U-\ov U)^{-n}\Delta_{q_1q_2q_3q_4}$ 
contains the piece
\eqn\leading{\eqalign{
{\cal I}_{w_T,w_U,w_{\ov T},w_{\ov U}}&={(T- \ov T)^{-m}\o (U-\ov U)^{n}}\cr
&\times \int {d^2\tau \o \tau^{1-r+s}_2}  
\sum_{(p_L,p_R)}\ov p_R^{q_1} p_R^{q_2} \ov p_L^{q_3} p_L^{q_4}\ 
e^{\pi i\tau|p_L|^2} e^{-\pi i\ov\tau|p_R|^2}\ \ov f_{k}(\ov \tau)\ ,\cr}}
next to subleading pieces, which are less harmonic and are due to IR--effects
originating from pinching vertex operators.
In general, those terms have different modular weights. 
${\cal I}_{w_T,w_U,w_{\ov T},w_{\ov U}}$ has modular weights 
\eqn\weights{\eqalign{
w_T&=m+\h(q_1-q_2-q_3+q_4)\cr
w_{\ov T}&=m+\h(q_2-q_1-q_4+q_3)\cr
w_U&=n+\h(q_2-q_1+q_4-q_3)\cr
w_{\ov U}&=n+\h(q_1-q_2+q_3-q_4)\cr}}
under $SL(2,{\IZ})_T\times SL(2,{\IZ})_U$, $(ad-bc=1)$:
\eqn\dulnar{\eqalign{
(p_L,\bar p_R)&\to\sqrt{cT+d \o c\bar T+d}\ (p_L,\bar p_R)\ \ \ ,\ \ \
T\to {{aT+b}\o {c T+d}}\ , \ U\to U\ , \cr
(p_L,p_R)&\to\sqrt{cU+d\o c\bar U+d}\ (p_L,p_R)\ \ \ ,\ \ \ 
T\to T\ ,\ U\to {{a U+b} \o {c U+d}\ }\ .}}
Poisson resummation on \setup\ leads to 
\eqn\todo{\eqalign{
\Delta_{q_1q_2q_3q_4}&:=
T_2{\p^N \o \p\Lambda_1^{q_1} \p\Lambda_2^{q_2} \p\Lambda_3^{q_3}
\p\Lambda_4^{q_4}}\int {d^2\tau \o \tau_2^{2-r+s}} 
\sum_{A\in M(2,\IZ)}\ e^{-2\pi i \ov T\det A}\ 
e^{{-\pi T_2\o\tau_2 U_2}|\lf({1\atop U}\ri)^tA\lf({\tau\atop -1}\ri)|^2}\cr
&\times \lf.e^{-{1\o \sqrt 2\tau_2}
(\Lambda_1\ov Q_R-\Lambda_2Q_R +\Lambda_3\ov Q_L-\Lambda_4 Q_L)}
\ \ov f_{k}(\ov \tau)\ri|_{\Lambda_i=0}\cr 
&=2^{-{N\o 2}}T_2(-1)^{q_2+q_4+N}\cr
&\times \int {d^2\tau \o \tau_2^{2+N-r+s}}
\sum_{A\in M(2,\IZ)} \ov Q_R^{q_1} Q_R^{q_2} \ov Q_L^{q_3} Q_L^{q_4} 
e^{-2\pi i \ov T\det A}
e^{{-\pi T_2\o\tau_2 U_2}|\lf({1\atop U}\ri)^t
A\lf({\tau\atop -1}\ri)|^2}\ \ov f_{k}(\ov \tau).\cr}} 
Here the sum runs over all integer $2\times 2$ matrices 
$A=\lf({n_1\ -l_2\atop n_2\ l_1}\ri)\in M(2,\IZ)$.
Modular invariance enables us to use the orbit decomposition 
used in \DKLII, i.e. decomposing the set of all
matrices $A$ into orbits of $SL(2,\IZ)$:
\eqn\orbits{\eqalign{
I_0&:\ A=\lf( {0 \atop 0}\ {0\atop 0}\ri)\ , \cr
I_1&:\ A=\pm\lf( {k \atop 0}\ {j\atop p}\ri)\ \ ,\ \ 0\leq j<k\ ,\
p\neq 0\ , \cr
I_2&:\ A=\lf( {0 \atop 0}\ {j\atop p}\ri)\ \ ,\ \ (j,p) \neq
(0,0)\ .}}
Since we excluded the case $(q_1,q_2,q_3,q_4)=(0,0,0,0)$, 
$I_0$ does not give any contribution.
The integral to be done for the $I_1$ orbit is:
\eqn\integral{
\int_0^\infty {d\tau_2\o \tau_2^{3/2}} e^{-a\tau_2-{b\o \tau_2}}=
\sqrt{\pi\o b}e^{-2\sqrt{ab}}\ .}  
After introducing
\eqn\short{\eqalign{
b&=p^2-
{i(\Lambda_1+\Lambda_2+\Lambda_3+\Lambda_4)p\o \pi \sqrt{2T_2U_2}}
-{1\o 8}{(\Lambda_1-\Lambda_2+\Lambda_3-\Lambda_4)^2\o \pi^2T_2U_2}\cr
\varphi&=p(kT_1+lU_1)\cr
&+{1\o 2\pi\sqrt{2T_2 U_2}}\lf[(kT_2-lU_2)\Lambda_1
+(-kT_2+lU_2)\Lambda_2+(-kT_2-lU_2)\Lambda_3+(kT_2+lU_2)\Lambda_4\ri]\ ,\cr}}
and the functions
\eqn\Ione{\eqalign{
I_1(\alpha,\beta)&=
{2 \o \sqrt{\beta b}}e^{-2\pi (kT_2+lU_2)\sqrt{\alpha\beta b}}
\ e^{-2\pi i\varphi}\ ,\cr
I_2(\alpha,\beta)&={1\o\sqrt{\beta b}}
e^{-2\pi lU_2\sqrt{\alpha\beta b}}
(e^{2\pi i\varphi}+e^{-2\pi i\varphi})\ .\cr }}
we obtain the closed expressions for $I_1$ and $I_2$ in the 
chamber $T_2>U_2$:
\eqn\Itwo{\eqalign{
I_1&={\p^N \o \p\Lambda_1^{q_1} \p\Lambda_2^{q_2}
\p\Lambda_3^{q_3} \p\Lambda_4^{q_4}}\times\cr
&\times\sum_{{k>0 \atop l \in \IZ}}\sum_{p\neq 0}c(kl)
\lf[{T_2U_2 \o \pi(kT_2+l U_2)^2}\ri]^r {1\o (\pi T_2 U_2 b)^s}
 {(-1)^{r+s}\p^{r+s}\o \p\alpha^r\p\beta^s} 
I_1(\alpha,\beta)\lf.\ri|_{{\alpha=1\atop \beta=1},\Lambda_i=0}\ ,\cr}}
\eqn\Itwoo{\eqalign{I_2&={\p^N \o \p\Lambda_1^{q_1} \p\Lambda_2^{q_2}
\p\Lambda_3^{q_3} \p\Lambda_4^{q_4}}\times\cr 
&\times\sum_{{k=0 \atop l>0}}\sum_{p\neq 0}c(0)
\lf[{T_2 U_2\o\pi(lU_2)^2}\ri]^r{1\o (\pi T_2 U_2 b)^s} 
{(-1)^{r+s}\p^{r+s}\o \p\alpha^r\p\beta^s}
I_2(\alpha,\beta)\lf.\ri|_{{\alpha=1\atop \beta=1}}\cr
&+c(0)sgn(s-r)^{s-r}(2\pi T_2 U_2)^{r-s}[1\cdot 3\cdot 5\cdot \ldots \cdot
(2|r-s|-1)]^{sgn(s-r)}\sum_{p\neq 0}{1\o b^{\h+s-r}}\cr
&+c(0)T_2^{r-s}{U_2^{1+s-r}\o \pi^{1+s-r}}(-1)^{r+s}{d^{s-r}\o d\xi^{s-r}}
\sum_{j\neq 0} {1\o j^2+j{1\o \pi}\sqrt{U_2\o
2T_2}(\Lambda_2+\Lambda_4-\Lambda_1-\Lambda_3)+ \xi}
\lf.\ri|_{{\xi=0\atop\Lambda_i=0}}\ .\cr}}
Generically, the second sum  leads to $\sum\limits_{p>0}p^{-a}=\zeta(a)$ for 
$a=1+2(s-r)+N\neq 1$. For $a=1$ one has to replace this term with a 
regularized sum, which is derived in the appendix.
In total, we obtain for \setup:
\eqn\total{
\Delta_{q_1q_2q_3q_4}=I_1+I_2^{reg.}\ .}

Let us briefly discuss the case $N:=q_1\neq 0,{\rm even}\ ,q_2,q_3,q_4=0$
\ ,\ $s=r=0$, for which
$I_1,I_2$ can be further simplified:
\eqn\simplify{\eqalign{
\Delta_N&=
{\p^N\o\p\Lambda^N} \times\cr
\times&\sum_{{k>0 \atop l \in \IZ}}\sum_{p\neq 0}c(kl)
{2 \o |p-{i \Lambda\o 2\pi \sqrt{2T_2U_2}}|} 
e^{-2\pi (kT_2+lU_2)|p-{i \Lambda\o 2\pi \sqrt{2T_2U_2} }|}\ 
e^{-2\pi i  p(kT_1+lU_1)-{i\Lambda\o \sqrt{2T_2 U_2}}(kT_2-lU_2)}\cr
+&\sum_{l>0}c(0)\sum_{p\neq 0}
       e^{-2\pi lU_2|p-{i\Lambda\o 2\pi\sqrt{2 T_2 U_2}}|}
\bigg(e^{ 2\pi i lpU_1}e^{-\Lambda il  \sqrt{U_2\o 2 T_2}}+
 e^{-2\pi i lpU_1}e^{ \Lambda il  \sqrt{U_2\o 2 T_2}}\bigg)\cr
+&2c(0)\zeta(N+1)N!\lf({i\o 2\pi \sqrt{2T_2 U_2}}\ri)^N
+2^{\h N+2}c(0)U_2\lf({U_2\o T_2}\ri)^{N\o 2}
{\pi\o (N+1)(N+2)}B_{{N\o 2}+1}\lf.\ri|_{\Lambda=0}\ .\cr}} 

Now let us come back to eq. \intFV.
Unfolding the integration \orbits\ enables one to identify these contributions
to tree-level, higher perturbative and
non-perturbative corrections on the type I side.
Let us briefly discuss a case, for which
$I_1,I_2$ can be  simplified further:
$q_1,q_2\neq 0;\ v,w=0$\ ,$s=r=0$ with $q\equiv q_2$.
The non-degenrate orbit  gives:
\eqn\intFV{
\Delta_{F_1^{4-q} F_2^q}^{\rm non-deg}=
{\partial^{4-q}\o\partial {\La_1}^{4-q}}{\partial^{q}
\o\partial {\La_2}^{q}}
\sum_{{0\le j<k\atop p\neq 0}}{1\o k\sqrt{b(\La_1,\La_2)}} 
e^{-2\pi i  k p {\tilde T}(\La_1,\La_2)}
{\cal A}[{\tilde U}(\La_1,\La_2)]\ ,}
where 
\eqn\btu{\eqalign{
\tilde T(\La_1,\La_2)&= T_1-{i\o p} T_2
\Big[\sqrt{b(\La_1,\La_2)}+{i(\La_1-\La_2)\o 2\pi\sqrt{2 T_2 U_2}}\Big]\cr 
\tilde U(\La_1,\La_2)&={1\o k}
\Big(j+p U_1-i U_2\Big[\sqrt{b(\La_1,\La_2)}+{i(\La_1-\La_2)\o 2\pi\sqrt{2
T_2 U_2}}\Big]\Big)\ .}}
The worldsheet instanton corrections
$\Delta^{\rm non-deg}$ are exponentially suppressed.
From the point of view of heterotic--type I duality it is 
convenient to rewrite this expression in terms of Hecke operators 
\BFKOV\ko:\foot{The Hecke operator acts on a modular form
$\Phi_w$ of weight $w$ as $H_N[\Phi_w](U)=\br
{1\o N^{1-w}}\sum_{k,p>0\atop kp=N}\sum_{0\le j<k} k^{-w}\Phi_w(\Uc)$ with
the $D$--brane complex structure $\Uc={j+p U\o k}$ \Serre.}
\eqn\instFV{\eqalign{
\Delta_{F_1^{4-q} F_2^q}^{\rm non-deg}&={\pi^4\o 2}
 \sum_{N=1}^\infty\bigg[\Big({T_2\o U_2}\Big)^{2-q}
\Big(D_{\Tc}^{4-q}q_{\Tc}\Big) 
N^{4-q} H_N[D^q_U\Ac_{-4}](U)\cr
&+\Big({U_2\o T_2}\Big)^{2-q}\ D_{\Tc}^q\bar q_{\bar \Tc} \ {1\o N^{4-q}}
\ H_N[{\bar D}^{4-q}_{\bar U}{\bar \Ac}_{-4}](\bar
U)\bigg]\ ,}}
where $q_{\Tc}=e^{2\pi i\Tc}$ with $\Tc=N T$ 
and covariant derivatives acting on $q_{\Tc}$ start with 
$D_{-4}^{\Tc}$. The structure of covariant derivatives in
the above expression matches with the one found in  \RES.

The degenerate orbit gives 
\eqn\degTTTT{
\Delta^{\rm deg}_{F_1^{4-q} F_2^q}={c_0\o\pi}\Gamma(5){U_2^3\o T_2^3}
\sum_{(j,p)\neq (0,0)}{(j-p U)^{4-q}(j-p \bar U)^{q} \o|j-p U|^{10}}\ .}
where $c_0\equiv c(0)=504$ is the constant coefficient 
of the light--cone Ramond partition function
${{\bar E}_4^2\o{\bar \eta}^{24}}=\sum_n c(n) \bar q^n$. 
The sum is taken over winding modes $j,p$. 
The case $n_1=n_2=0$ in \narimp\ corresponds to vanishing winding numbers. 
Only Kaluza-Klein momenta contribute to $\Delta^{\rm deg}$, 
which will be identified with type I perturbative corrections in the
next section. For $q=2$ the result can be expressed in terms of
the generalized Eisenstein function 
$E(U,3)=U_2^3\zeta(6)\sum_{(j.p)\neq 0}{1\o |j-p U|^6}$ \Terras.

\subsec{$R^4 T^{4g-4}$ couplings}

The heterotic effective action compacified on $K3\times T^2$ contains higher
derivative couplings $F_g \Tr R^2 T^{2g-2}$, 
where $T_{\mu\nu}$ is the field 
strength of the graviphoton. These couplings appear at one
loop \agnt\ and only depend on the vector multiplet scalars.
They do not depend on the $K3$ moduli.\foot{However, they depend 
on the chosen gauge bundle. See e.g.: \stieberg.} 
 Therefore the moduli--dependence
of these couplings should survive in the limit of large $K3$, i.e. 
in the decompactification to $d=8$. 
In fact, we will show in this section, that the one--loop corrections
to the couplings  $F_g  \Tr R^4 T^{4g-4}$ and $F_g (\Tr R^2)^2 T^{4g-4}$
share the same moduli dependence. 
They are derived from a one--loop amplitude 
with four graviton and $(4g-4)$ graviphoton vertex operators. 
Since the $CP$ odd amplitude is related
to the $CP$ even amplitude by supersymmetry, which results in  
the same coupling, it is sufficient to discuss the $CP$ even case only. 
This part is highly constrained by $U(1)$ charge conservation of the 
internal $c=3$ conformal field theory.
Essentially this means, that only the bosonic part of the graviphoton vertex
operators gives non--vanishing contributions.
For the $CP$-even amplitude we use \grvert\ as graviton vertex operator in  
zero ghost picture and the following graviphoton vertex operator 
in background gauge:
\eqn\gphoton{
V_{gph}^{(0)}=-\hf T_{\mu\nu} \bar \p X^\mu[X^\nu \e_I\p X^I-{1\o 4}
\e_I\gamma^{\nu I}_{ab} S^a S^b]\ . }
The eight fermionic zero modes are soaked up by the graviton vertex operator
and we get for the amplitude
\eqn\cpeven{\eqalign{
\Ac&=\Big\la \int \prod_{i=1}^4 d^2z_i V_{g,i}^{(0)}(z_i)\int 
\prod_{j=1}^{4g-4}d^2z_j 
V_{gph,j}^{(0)}(z_j) \Big\ra\cr
&=\Big({1\o 2} \Big)^{4g} t^{\nu_1\sigma_1\nu_2\sigma_2\nu_3
\sigma_3\nu_4\sigma_4}
\prod_{i=1}^4 R_{\mu_i\rho_i\nu_i\sigma_i}\prod_{i=5}^{4g}
T_{\mu_i\rho_i} \cr
&\times\int {d^2\tau\o \tau_2^5} {\bar E_4^2\o \bar{\eta}^{24}}
\sum_{p_L, p_R} \Big({p_L\o \sqrt{2 T_2 U_2}}\Big)^{4g-4}
q^{\h p_L^2} {\bar q}^{\h p_R^2}
\Big\la \int \prod_{i=1}^{4g} d^2z_i \bar\p X^{\mu_i}(z_i) X^{\rho_i}(z_i)
\Big\ra\ .}}
For the bosonic correlation function we introduce the generating functional 
\eqn\gen{
G(\tilde\lambda_1,\tilde\lambda_2, T, U)=\sum_{g=1} {1\o {2g!}^2}\Big(
{\tilde\lambda_1\tilde\lambda_2\o\tau_2}\Big)^{2g-2}
\la\prod_{i=1}^{2g}
\bar\p X^{\mu_i} X^{\nu_i}\ra^2=
\tau_2^4 {d^4\o d{\tilde\lambda_1}^2d{\tilde\lambda_2}^2}
\la e^{-S_0+\tilde\lambda_1 S}\ra \la e^{-S_0+\tilde\lambda_2 S}\ra}
with $\tilde\lambda_i={\lambda_i \tau_2 p_L\o \sqrt{2 T_2 U_2}}$
and \foot{In analogy to \agnt\ we evaluate the determinant 
by passing to a complex basis for the eight dimensional Euclidean space:
$Z^1={1\o\sqrt{2}}(X^1-i X^2)$, $Z^2={1\o\sqrt{2}}(X^0-i X^3)$,   
$Z^3={1\o\sqrt{2}}(X^5-i X^6)$,   $Z^4={1\o\sqrt{2}}(X^4-i X^7)$.}     
$S_0-\tilde\lambda S=
{1\o \pi}\int d^2 z \Big(\p X^\mu \bar \p X_\mu-{\tilde\lambda\o \tau_2}
X^\mu\bar\p X_\mu\Big)$ and
\eqn\determ{
\la e^{-S_0+\lambda S}\ra={|\eta|^4\o
(\det'\Delta)^2}=
\Big[{\lambda\theta_1'(0,\bar\tau)\o \theta_1(\tilde\lambda,\bar\tau)}\Big]^2
e^{-\pi\tilde\lambda^2\o \tau_2}\ ,}
where $(\det'\Delta)^{d/2}=|\eta|^{d} 
\Big[{\theta_1(\tilde\lambda,\bar\tau)\o 
{\bar \eta}^3\tilde\lambda 2\pi}\Big]^{d/2} 
e^{d \pi\tilde\lambda^2\o 4\tau_2}$ 
is the determinant for $d$ scalar fields on a worldsheet torus
which is calculated using $\zeta$--function
regularization \agnt. We introduced the dependence on $\tilde\lambda_1$
and $\tilde\lambda_2$ which keeps track of the space--time kinematics
of the tensors $R_{\mu\nu}$ and $T_{\mu\nu}$ w.r.t. complex bosons $Z^1,Z^2$
and $Z^3,Z^4$, respectively.
We may summarize all couplings $F_g$ by the generating function
\eqn\flam{\eqalign{
F(\tilde\lambda_1,\tilde\lambda_2,T,U)&=\sum_g 
F_g\ (\lambda_1\lambda_2)^{2g-2}= 
\int {d^2\tau\o \tau_2} {\bar{E}_4^2\o \bar{\eta}^{24}}\sum_{p_L, p_R}
q^{\h p_L^2}\bar{q}^{\h p_R^2} \cr
&\times {d^4\o d\tilde\lambda_1^2d\tilde\lambda_2^2} 
\lf[{2\pi i\tilde\lambda_1\ov\eta^3\o \theta_1(\tilde\lambda_1,\ov \tau)}
e^{-{\pi\tilde\lambda_1^2\o 2\tau_2}}
\ {2\pi i\tilde\lambda_2\ov\eta^3\o \theta_1(\tilde\lambda_2,\ov \tau)}
e^{-{\pi\tilde\lambda_2^2\o 2\tau_2}}\ri]^2\ .\cr}} 
Using the identity 
${\tilde \lambda\theta_1'\o \theta_1(\tilde\lambda,\bar\tau)}
e^{-{\pi\tilde\lambda^2\o 2\tau_2}}=
\exp\Big[\sum_k{1\o 2k}\tilde\lambda^{2k} \hat{G}_{2k}\Big]$ 
the above expression can be compared with the gravitational part of
the elliptic genus \ellg, replacing ${R\o 4\pi^2}$ by $\tilde\lambda$.

We use a somewhat different way than the one in \MM\ to integrate \flam.
In fact, similar as in \LSW,  it proves to be useful
to promote the Jacobi--form
$\Phi_{-2}(\tau,\z)={1\o \theta_1(z_1,\tau)^2\theta_1(z_2,\tau)^2}$ 
of characteristics $(w,m)=(-2,-1)$ to the almost holomorphic function 
\eqn\tildejac{
\tilde\Phi_{-2}(\tau,\z)={1\o \theta_1(z_1,\tau)^2\theta_1(z_2,\tau)^2}
e^{\pi {\z^2 \o \tau_2}m}\ \ ,\ \ m=-1\ ,}
with the well--behaved modular behaviour 
$\tilde\Phi_{-2}\lf({a\tau +b \o c\tau+d},{\z\o c\tau+d}\ri)=
(c\tau+d)^{-2}\tilde\Phi_{-2}(\tau,\z)$.
Thus all the non--holomorphic parts of the genus are captured
by $\tilde\Phi_{-2}$.
We treat the $p_L^{4g-4}$--lattice vector insertions as before in eq. 
\setup\ and obtain for $F_g$ the closed form
\eqn\TODO{\eqalign{
F_g&={1\o (2g)!^2}{1\o(2T_2U_2)^{2g-2}}{d^{4g-4}\o d\Lambda^{4g-4}}
{d^{4g}\o d\ov z_1^{2g}d\ov z_2^{2g}}\times\cr
&\times\int{d^2\tau\o\tau_2}
\sum_{(p_L,p_R)} q^{\h|p_L|^2}\ov q^{\h|p_R|^2}
e^{\tau_2\Lambda p_L}{\ov E_4^2\o \ov \eta^{24}}
(2\pi i\ov \eta^3)^4 \ov z_1^2\ov z_2^2
\tilde \Phi_{-2}(\ov \tau,\ov \z)\lf.\ri|_{{\ov \z=0\atop \Lambda=0}}\ ,\cr}}
in which the $\tau$--integration becomes straightforward.
We need the coefficients $c(n;\b)$ appearing in the power 
series\foot{The functions ${\cal P}_{2b}(G_2,\ldots,G_{2b})$ 
have been introduced in \MM.
E.g.: ${\cal P}_{0}=-1,\ {\cal P}_{2}=-G_2,\ {\cal P}_{4}=-\h(G_2^2+G_4)$,
${\cal P}_{6}=-{1\o 3}G_6-\h G_2G_4-{1\o 6}G_2^3$, 
${\cal P}_{8}=-{3\o 8}G_4^2-{1\o 3}G_2G_6-{1\o 4}G_4G_2^2-{1\o 24}G_2^4$.}
$(y_k=e^{2\pi i z_k})$:
\eqn\expansion{
{E_4^2\o \eta^{24}}
(2\pi i \eta^3)^4 z_1^2z_2^2\Phi_{-2}(\tau,z)={E_4^2\o \eta^{24}}
\sum_{b_1,b_2=0}^\infty z_1^{2b_1}z_2^{2b_2}{\cal P}_{2b_1} {\cal P}_{2b_2} 
=\sum_{b_1,b_2=0}^\infty\sum_{n \geq -1} c(n;\b) q^n z_1^{2b_1}z_2^{2b_2}\ .}
Only $F_1$ receives a contribution from the trivial orbit of \orbits, namely
\eqn\trivial{
I_0(g)=\cases{       {16\o 3}\pi^5T_2\ \ : & $g=1$\cr 
                         0           \ \ : & $g\neq 1$\ .}}
Proceeding with the orbits \orbits\ we obtain for the non--degenerate 
orbit\foot{We already include the $k=0, l>0$ contribution coming from
the degenerate orbit $I_2(g)$.} (in the chamber $T_2>U_2$):
\eqn\nd{\eqalign{
I_1(g)&={1\o (2g)!^2}{1\o(2T_2U_2)^{2g-2}}{d^{4g-4}\o d\Lambda^{4g-4}}
{d^{4g}\o d z_1^{2g}d z_2^{2g}}\sum_{\b}\sum_{(k,l)>0}c(kl;\b)
z_1^{2b_1}z_2^{2b_2}\cr
&\times\sum_{p\neq 0}{2\o\sqrt{p^2-{m\z^2\o T_2U_2}}}
e^{-2\pi\sqrt{p^2-{m\z^2\o T_2U_2}}|kT_2+lU_2+{i\Lambda \sqrt{T_2U_2} 
\o 2\pi\sqrt{2}}|}
e^{-2\pi ip[(kT_1+lU_1)-{\Lambda \sqrt{T_2U_2} \o 2\pi\sqrt{2}}]}
\lf.\ri|_{{\Lambda=0\atop \z=0}}\cr
&={1\o (2g)!^2}{(-1)^{2g-2}\o 2^{4g-5}}{d^{4g}\o d z_1^{2g}d z_2^{2g}}
\sum_b\sum_{(k,l)>0}c(kl;\b)z_1^{2b_1}z_2^{2b_2}\cr
&\times\sum_{p\neq 0}{\Big(p-\sqrt{p^2-{m\z^2\o T_2U_2}}\Big)^{4g-4}\o 
\sqrt{p^2-{m\z^2\o T_2U_2}}} e^{-2\pi\sqrt{p^2-{m\z^2\o
T_2U_2}}(kT_2+lU_2)}e^{-2\pi i p(kT_1+lU_1)}\lf.\ri|_{\z=0}.\cr}}
We have checked, that after some work, \nd\ can be written for 
$z_1=z_2$ in the form
eq. (4.40) of \MM. In \nd, the sum $p<0$ gives rise to $T,U$ dependent 
polylogarithmic contributions, whereas $p>0$ leads to 
$\ov T,\ov U$ dependent pieces, which vanish for $g\geq 3$. 
For the degnerate orbit we find:
\eqn\rd{\eqalign{
I_2(g)&={1\o (2g)!^2}
{d^{4g}\o d z_1^{2g}d z_2^{2g}}\sum_{\b} 2^{4-4g}
c(0;\b)z_1^{2b_1}z_2^{2b_2}\cr&\times\lf\{(-1)^{2g-2} \sum_{p\neq 0} 
{\Big(p-\sqrt{p^2-{m\z^2\o T_2U_2}}\Big)^{4g-4}\o 
\sqrt{p^2-{m\z^2\o T_2U_2}}}+{2U_2^{5-4g}\o \pi}
\sum_{j=1}^\infty {j^{4g-4}
\o j^2-m\z^2{U_2\o T_2}}\ri\}\lf.\ri|_{z=0}.\cr}} 
For $g=1$ the first sum in the bracket of \rd\ is divergent 
if none of the $z_i$--derivatives act on it. In general, this divergence
occurs when $(4g-4)$ $z_i$--derivatives act on the sum
over $p$. In these cases the sum has to be replaced by a regularized 
expression, which we present in the appendix. 
With that in mind, we get in total for \TODO: 
\eqn\alt{
F_g=I_0(g)+I_1(g)+I^{reg.}_2(g)\ .}
We want to extract the holomorphic piece of $F_g$,
which has a topological interpretation on the 
dual IIB side, described by $F$--theory on $K3$. 
The latter is described by an N=4 superconformal field theory
and $4g-4$ graviphoton insertions lead to topological invariant
correlators at $g$--loop order \bvi.
The holomorphic piece $F_g^{holom.}$ 
appears from \nd\ and \rd\ in the limit $\ov
T,\ov U\rightarrow i\infty$, i.e. when all
$z$--derivatives act on the genus $\sum_{\b} c(n;\b)z_1^{2b_1}z_2^{2b_2}$:
\eqn\final{
F^{holom.}_{g\geq 2}(T,U)=
\zeta(5-4g)c(0;2g,2g)+2\sum_{(k,l)>0} c(kl;2g,2g)\ \Li_{5-4g}
\big[{q_T}^k {q_U}^l\big]\ .}
The harmonic part of $F_1$ is given by:
\eqn\FONE{
F_1^{harm.}(T,U,\ov T,\ov U)=-{24i \pi^5\o 9}T-
{76i \pi^5\o 9}U+
2\sum_{(k,l)>0}\ c(kl;2,2)\Li_1\big[{q_T}^k {q_U}^l\big]+hc.\ ,}
whereas the case $g=0$ is related to \prep.
After \expansion, the coefficients $c(kl;2g,2g)$ refer to
$\cP_0^2E_4^2\eta^{-24}=E_4^2\eta^{-24}, 
\cP_2^2E_4^2\eta^{-24}=G_2^2E_4^2\eta^{-24}$ and 
$\cP_4^2E_4^2\eta^{-24}={1\o 4}(G_4+G_2^2)^2E_4^2\eta^{-24}$
for $g=0,1$ and $g=2$, respectively.
Clearly, for $g=0$ we just get the holomorphic prepotential ${\cal G}$
\prep, which counts genus $0$ curves of the underlying $K3$ \LS.
The $g=1$ case gives the correction to $(\tr R^2)^2$, whereas
the $g=2$ case corresponds to $(\tr R^2)^2 T^4$.
Different distributions of $z_1,z_2$--derivatives in \TODO\ lead to different
Lorentz kinematics. In particular it is possible to find a combination
which gives a relation 
\eqn\relat{
F_g^{holom.} \leftrightarrow 
\lf({d \o dq} G_2(q)\ri)^g {q \o \eta(q)^{24}} E_4(q)^2} 
to the counting formula \BL. The latter counts genus $g$ curves with 
$g$ nodes on the $K3$ of the dual $F$--theory compactification.
A similar correspondence has been recently established in four dimensions
in counting curves of the $K3$ fiber of a Calabi--Yau manifold \MM.

The amplitudes $F_g$ have no obvious relation to the $R^4H^{4g-4}$
IIB couplings of \bvii. In particular, the latter contain the set of $(p,q)$
$D$--string contributions, which does not exist in our vacua.
Moreover the couplings \final\ behave like 
$F_g\rightarrow\zeta(5-4g)c(0;2g,2g)$
in the decompactification limit $T\rightarrow i\infty$
to ten dimensions in contrast to the IIB coupings $R^4H^{4g-4}$, 
which give the famous $D$--instanton sum.

\newsec{Type\ I in eight dimensions}

In this section we discuss two type I models $(A)$ and $(B)$. Model 
$(A)$ has $E_8\times E_8\times U(1)^2$ 
and model $(B)$\ $SO(8)^4\times U(1)^2$ gauge group.
For model $(A)$ we calculate gauge
threshold corrections w.r.t. to the $U(1)$ factors,
while for model $(B)$ we derive the corrections w.r.t. to the $SO(8)$.
Furthermore, we discuss type I $R^2$ corrections.

\subsec{$F^4$ type I couplings in model $(A)$}

The dilaton dependence for surfaces with Euler number $\chi=2-2g-B-C$,
is $\lambda_I^{-\chi}$ where $g$ is the genus, $B$ and $C$ the number 
of boundaries and crosscaps, respectively. 
Using the duality map \dualmap\
we verify that the degenerate orbit of the heterotic threshold
corrections \degTTTT
\eqn\hetdeg{
S^{\rm het}_{\rm deg}={V^{(8)} T_2\o 2^8\pi^4} t_8 F_1^{4-q}F_2^q
\Delta^{\rm deg}_{F_1^{4-q} F_2^q}(T_2,U)
\leftrightarrow S^{\rm I}_{\rm 1-loop}\ ,}
appear at one loop in the type I effective action. 
The term $V^{(8)} t_8$ is invariant under this transformation,
and the factor $e^{2\Phi^I}$ which arises  from
${T_2\o T_2^3}$ is canceled by  $G_{\rm int}^{-2}$ contracting the 
internal indices of the gauge
kinetic term. Thus on the type I side we get the corresponding 
one loop ($\lambda_I^0$) coupling.

Now let us directly compute \hetdeg\ in type I string theory.
One loop open string amplitudes consist of
summing over oriented and unoriented surfaces 
with and without boundaries like the
torus ($\Tc$) and Klein bottle ($\Kc$) for the closed string sector and
the annulus ($\Ac$) and M{\"o}bius  strip ($\Mc$) for the open string sector,
which have Euler number $\chi=0$.

The vertex operator for Abelian gauge fields coincides with
the one of type IIB theory:
\eqn\vert{
V_{\rm gauge}=G_{\mu I} :\Big(\bar \partial X^I-{1\o 4}
k_\sigma \ov S^a \gamma^{I\sigma}_{a,b}\bar S^b\Big)
\Big(\partial X^\mu-{1\o 4}k_\nu S^a\gamma^{\nu\mu}_{ab}S^b\Big) e^{i k
  X}:\ .}
There is no contribution from the torus diagram $\Tc$ since the sixteen
fermionic zero modes cannot be saturated at the level of four
derivative terms. 
The remaining amplitudes can be written as:
\eqn\oampl{\eqalign{
S_{\rm 1-loop}^I&={1\o 2} V^{(8)}\sum_{\sigma=\Kc,\Ac,\Mc}\rho_{\sigma}
\int_0^\infty {dt\o t}{1\o (2\pi^2 t)^4}
\Big(\sum_{p\in \Gamma_2}e^{-\pi t |p|^2/2}\Big)\  Z(\tau_{\sigma})\cr
&\times \int\prod_{i=1}^4 d^2 z_i\la\prod_{i=1}^4 {V}_{{\rm gauge},i}\ra}}
with relative weights $\rho_{\Kc}=1$, $\rho_{\Ac}=N^2$ and
$\rho_{\Mc}=-N$, and $N$ is the Chan-Paton charge which is
$N=32$ for $SO(32)$.
The factor $(2\pi^2 t)^{-4}$ arises from momentum integration and
$V^{(8)}$ is the uncompactified volume in type I units.
The open string oscillator sum is
$Z(\tau_{\sigma})={1\o \eta^{12}(\tau_{\sigma})}\sum_{\alpha=2,3,4}\h
s_\alpha\theta_\alpha^4(0,\tau_{\sigma})$
with GSO projection signs 
$s_3=-s_2=-s_4=1$ and modular parameters
$\tau_{\Ac}={it\o 2}, \tau_{\Mc}={it+1\o 2}, \tau_{\Kc}={2it}$.
The perturbative duality group for the open string  is reduced to
$SL(2,{\bf Z})_U$ and 
the torus partition function $\Gamma_2$ is now
restricted to the Kaluza-Klein momenta $m_1$ and $m_2$ and
$p^2=p_I G^{IJ} p_J={1\o 2 U_2 \sqrt{G}}|m_1+m_2 U|^2$.

Contraction of the leftmoving fermions $S^a\gamma^{\nu\mu}_{ab} S^b$
contributes  four derivatives to the amplitude and  
using a Riemann identity resulting in
\eqn\rid{
Z(t)G_F^4(t)=-{\pi^4\o 2}\ .}
one finds after Poisson resummation:
\eqn\typeIthr{\eqalign{
S_{\rm 1-loop}^I&={V^{(8)} T_2\o 2^8}  t_8 F_1^{4-q} F_2^q
\int {dt\o t^6}\sum_{j,p\neq 0}
e^{{-\pi T_2\o t U_2}|j-p U|^2} {T_2^2\o U_2^2}\cr
& \times (j-p U)^{4-q} (j-p \bar U)^q {1\o 2}\Big(N^2-N+2^4\Big)}}
where $\hf (N^2-N+2^4)=c_0$.
Integration over $t$ thus reproduces the corresponding heterotic coupling
$\Delta^{\rm deg}_{F_1^{4-q} F_2^q}(T_2,U)$.
\subsec{$R^4T^{4g-4}$ type I couplings}
The relevant type I one-loop amplitude involves four gravitons and
$(4g-4)$ graviphotons whose vertex operators are: 
\eqn\vgvgr{\eqalign{
V_{\rm grav}&=\e_{\mu\nu}\ :
\Big(\bar {\p} X^{\nu}-{1\o 4} \bar S^a\gamma^{\nu\rho}_{ab}
\bar S^{b} k_{\rho}\Big) 
\Big({\p} X^{\mu}-{1\o 4} S^{a}\gamma^{\mu\sigma} S^b  k_{\sigma}\Big)
e^{i k X}:\cr
V_{\rm gph}&=G_{\mu I} :\Big(\bar \partial X^\mu-{1\o 4}
k_\sigma \ov S^a \gamma^{\mu\sigma}_{a,b}\bar S^b\Big)
\Big(\partial X^I-{1\o 4}k_\nu S^a\gamma^{\nu I}_{ab}S^b\Big) e^{i k X}:\ .}}
In analogy to \apt\  we can immediately write down the expression
for the generating functional:
\eqn\typeIgen{
F_g^I(\lambda)=\sum_{\sigma=\Tc,\Ac,\Mc,\Kc}\rho_{\sigma}
\int_0^\infty {dt\o t}
\sum_{p\in \Gamma_2}e^{-\pi t |p|^2/2}\  
{d^4\o d\tilde\lambda^4} \la 
e^{S_0+\tilde \lambda S_{\rm inst}}\ra_{\sigma}\ ,}
with 
$\la e^{S_0+\tilde \lambda S_{\rm inst}}\ra_{\Ac}
=\la e^{S_0+\tilde \lambda S_{\rm inst}}\ra_{\Mc}
=\Big({\pi\tilde\lambda\o \sin\pi\tilde\lambda}\Big)^4$
and 
$\la e^{S_0+\tilde \lambda S_{\rm inst}}\ra_{\Kc}=
\Big({2\pi\tilde\lambda\o \tan\pi\tilde\lambda}\Big)^4$.
The torus amplitude only contributes to $F_1^I$, which   
yields $\ln U_2|\eta(U)|^4$ \PK\ from summing up even-even and 
odd-odd spin structures. 

The final expression is:
\eqn\ffinal{\eqalign{
F_g^I(\lambda)&=\int_0^\infty {dt\o t}
\sum_{p\in \Gamma_2}e^{-\pi t |p|^2/2}\ \bigg[{N^2-N+2^4\o 2} 
{d^4\o d\tilde\lambda^4} 
\Big({\pi\tilde\lambda\o \sin\pi\tilde\lambda}\Big)^4\cr
&+{d^4\o d\tilde\lambda^4}\Big((2\pi\tilde\lambda)^4
-2{(2\pi\tilde\lambda)^4\o(\sin\pi\tilde\lambda)^2}\Big)\bigg]
+F_1^{\rm torus}\ ,}}
which can be identified with the degenrate orbit of \flam for 
$\tilde\lambda_1=\tilde\lambda_2=\lambda$.
\subsec{$SO(8)^4$ type I threshold corrections in model $(B)$}

In this section we discuss the orientifold example of Sen \sen.
For this model heterotic/F-theory duality can be checked explicitly \LS. 
The (discrete) Wilson lines $a_1^I=\hf(0^4,0^4,1^4,1^4)$ and 
$a_2^I=\hf (0^4,1^4,0^4,1^4)$ break the gauge group to $SO(8)^4$.
For this model the internal Abelian gauge group cannot be 
enhanced  \pw\ and the underlying prepotenial is trivial. 
I.e. in this case the corresponding
one--loop gauge couplings $\Delta_{F_1^{4-q}F_2^q}$ in \RES\ 
vanish identically \LS.

Since the $SO(8)^4$ arises from Chan--Paton factors with constant gauge
field background $a_I$, we apply the method developed in
\BP\BF\bk\ to calculate type I one-loop threshold corrections.
In this setup the expression for the one-loop
amplitude of the open string reads
\eqn\amplbg{\eqalign{
S_{\rm 1-loop}^I& ={i\o 2} V^{(8)}\sum_{\sigma=\Ac,\Mc;ij}\rho_{\sigma}  
\int {dt\o t} {1\o (2 \pi^2 t)^4}
\sum_{a_{ij}^{\sigma}+\Gamma_2} e^{-\pi t p^2/2} {1\o \eta^{12}({
\tau_\sigma})}
 {i\o 2} q_{ij}^\sigma B t\cr 
&{\theta_1'(0,\tau_{\sigma})\o \theta_1\Big({i\e_{\sigma} t\o 2},
\tau_\sigma\Big)}
 \sum_{\alpha}\hf s_{\alpha}\theta_{\alpha}\Big({i\e_{\sigma} t\o
2},\tau_\sigma\Big)
\theta^3_{\alpha}(0,\tau_{\sigma})\ ,}}
where $F=B Q$ is the background gauge field and $Q$ a generator of
the Cartan subalgebra and $q_i$ the corresponding charge of the open
string endpoint carrying Chan-Paton index i.
The non-linear function $\e_\sigma$ can be expanded as 
$\e_\sigma\simeq q_{\sigma} B+\Oc(B^3)$ with $q_{ij}^{\Ac}=(q_i+q_j)$ and 
$q^{\Mc}_{ij}=2 q_i$. 

Expanding the integrand to the order $\Oc(B^4)$ gives
\eqn\anmoeb{
S_{\rm 1-loop}^I=-\hf V^{(8)}\sum_{\sigma=\Ac,\Mc;ij} 
\rho_\sigma  \int {dt\o t}
\Big[\e_{\sigma}^4
\sum_{a_{ij}^\sigma+\Gamma_2}e^{-\pi t p_I G^{IJ} p_J/2}\Big]\ .}
After Poisson resummation and changing variables from 
direct channel to closed string transverse channel
which is $l=1/t$ for the annulus and $l=1/(4t)$ for the M\"obius strip,
one finds:
\eqn\acmc{\eqalign{
S^I_{\rm 1-loop}&=-{i\o 2\pi} V^{(8)} B^4
\sum_w{T_2\o w^I G_{IJ} w^J}\bigg[ \sum_{ij}
e^{2\pi i (a_i+a_j)_I w^I}(q_i+q_j)^4\cr
&-\sum_i e^{4\pi i a_{iI} w^I}(2 q_i)^4\bigg]\ .}}
Evaluating the sum leads to
\eqn\sectors{\eqalign{
S^I_{\rm 1-loop}&=i\pi V^{(8)}B^4\bigg[4\ln\Big[T_2 U_2|\eta(U)|^4\Big]
\sum_{i<j; (i,j)=({k},{k})}(q_i+q_j)^4\cr
&+2\sum_{k=2}^4 \ln\Big[T_2 U_2\Big|{ \theta_k(U)\o 2 \eta(U)}\Big|^2\Big]
\sum_{(i,j)=({\bf 1},{\bf k})}(q_i+q_j)^4 \bigg]}}
where $k={\bf 1},\ldots,{\bf 4}$ and
${\bf 1}=\{1,\ldots,4\}, {\bf 2}=\{5,\ldots,8\},{\bf
3}=\{9,\ldots,12\}, {\bf 4}=\{13,\ldots,16\}$.
We omitted some moduli independent constant which appears after 
regularization of the logarithmic divergence \DKLII.

In the T-dual type I' picture this example corresponds to
placing four seven branes at each of the
four fixed points. The above threshold
corrections to ${\rm Tr}F^2_{SO(8)_k} {\rm Tr} F^2_{SO(8)_{k'}}$
arise from open strings stretched 
between branes sitting on the same or different fixed points: 
e.g. for $k=k'$ we have
\eqn\acmcfin{
S_{\rm 1-loop}^I=  i\pi V^{(8)}
\ln\Big[T_2 U_2|\eta(U)|^4\Big] \  \Big(\Tr F_{SO(8)_{k}}^2\Big)^2\ .}
On the heterotic side they translate to the degenerate orbit of
the corresponding coupling.
\subsec{D--instanton Contribution}
Heterotic/type\ I duality maps winding modes of 
the heterotic string to winding modes of $D1$-strings.
Heterotic worldsheet instantons that
arise in the non-degenerate orbit 
$A=\Big({k\ j\atop 0\ p}\Big)$ in the one loop amplitude,
are 'dual' to BPS instantons which arise
after wrapping the euclidean worldsheet of $D1$ branes
on the spacetime $T^2$. 
Choosing a basis of two-cycles on $T^2$, all inequivalent
ways in which a $T^2$ can cover $N$-times  another $T^2$ are given
by transformations that can be written in the form of the matrix $A$
with $kp=N$ and $0\le j<k$ and $p>0$ and $N$ is the instanton number.    
The embedding of the $D1$-string
worldsheet into the target space torus is then given by $X_i=A_{ij}\sigma^j$.

The classical instanton saddle point 
is the exponent of
the  Born-Infeld action of the wrapped $D1$-brane \BFKOV\ko: 
\eqn\scl{
S_{BI}= {1\o \lambda_I}\int d^2\sigma \sqrt{\det 
(\Gc_I+\Fc)}
-i\int {\Bc}_I^R\ ,}
where ${\Gc}$ and ${\Bc}_I^R$ are the pull-backs of the
metric and the antisymmetric tensor and $\Fc$ the $U(1)$ 
gauge background of the open string.

Fluctuations around the classical instanton saddle point are described 
by the elliptic genus $\Ac(U)$ (and its descendents) for $N$ $D1$-branes
on $T^2$. Similarly we can take a single $D1$-brane 
wrapped $N$ times over $T^2$. This is realized by
the action of the Hecke operator on the elliptic genus 
$H_N[\Ac]({\cal U})$. The complex structure $U$ of the
space-time $T^2$ is modified according to $\Uc={j+p U\o k}$
with the conditions for $j,p,k$ as above, e.g.
the three inequivalent ways a torus can double cover another torus is
$2 U$, ${U\o 2}$ and ${\h+{U\o 2}}$.

Let us  consider  $\Delta_{R^4T^{4g-4}}^{\rm non-deg}\equiv I_1(g)$ from 
\nd\ ($\tilde\lambda_1=\tilde\lambda_2=\lambda$) as an example, 
which in type I variables becomes  
\eqn\ttuu{\eqalign{
\la R^4 T^{4g-4}\ra_{\rm inst}
&={1\o (4g)!}{(-1)^{2g-2}\o 2^{4g-5}}{d^{4g}\o {d z}^{4g}}
\sum {1\o T_2^{4g-3} k^{4g-2}}\cr 
\times &{\Big[\sqrt{\det\Gc}-
\sqrt{\det(\Gc+\Fc(z))}\Big]^{4g-4}\o 
\sqrt{\det(\Gc+\Fc(z))}} e^{-2\pi S_{BI}} \Ac(\tilde U(z))\ ,}}
where 
\eqn\boin{\eqalign{
\tilde U(z) & ={1\o k}\Big[j+p U_1-i p U_2
\sqrt{\det(\Gc+\Fc(z))}/\sqrt{\det\Gc}\Big]\cr 
\sqrt{\det(\Gc+\Fc(z))}&=T_2 k 
\sqrt{p^2-{m z^2\o T_2 U_2}}\cr
\sqrt{\det\Gc}&=T_2 k p\cr
{\Bc}_I^R&=k p T_1\ ,}}
and the skew symmetric eigenvalues of $\Fc$ are $f^2=-
{z^2k^2m T_2\o U_2}$.

In analogy to semiclassical instanton 
calculations the correlation
function in an instanton background is obtained
by saturating fermionic zero modes and integrating over the
moduli space of instantons. In this case
the instanton moduli space is provided 
by the heterotic matrix string model \lowe, which describes
a worldsheet $O(N)$ two dimensional gauge theory.
In the infrared limit this gauge theory flows
to a $(8,0)$ supersymmetric  $S_N\times {\bf Z}_2^N$ 
orbifold conformal field theory. The elliptic genus
for $S_N$ symmetric orbifolds is naturally described
by the action of the $N$'th Hecke operator on the elliptic genus \dvv. 
It is known from \DMVV \ and \gn\  that the elliptic genus 
for $S_N$ symmetric orbifolds is given by the action of the $N$'th 
Hecke operator on the elliptic genus.

\subsec{One-Loop corrections to $R^2$}

Let us first briefly review threshold corrections  
in four dimensional $N=2$ string vacua, which are torus
compactifications of six dimensional $N=1$ string vacua.
In heterotic string vacua the Einstein term
does not get any one loop corrections. This fact gives rise to a relation
between threshold gauge couplings and the K\"ahler metric \anton.
Since in type I
theory Newton's constant is corrected at one loop  \ab\antlect, 
the dilaton has to be redefined, which 
provides the one loop correction to the type I K\"ahler metric.

In eight dimensions this situation translates to the
$R_{\mu\nu\rho\sigma} R^{\mu\nu\rho\sigma}$ coupling.
There are no heterotic one-loop corrections, whereas in
type I theory we will find a non-vanishing correction.
The graviton vertex operator in the zero ghost picture is given in \vgvgr.
Let us start with a two point amplitude. Although the kinematic
structure of a two point amplitude vanishes
due to the on shell constraints we can still calculate the 
four derivative gravitational coupling.
Going to a three point amplitude which e.g. includes a modulus and
two gravitons will give a non-vanishing kinematic structure and 
produce a derivative on the two point  coupling 
with respect to the modulus. 

The torus amplitude for a two point graviton vertex insertions vanishes
since the fermionic zero modes cannot be saturated.
But we can still get contributions from $\Kc$, $\Ac$ and $\Mc$.
We want to extract the order $\Oc(k^4)$ term of the amplitude, which
will only  produce a non-vanishing result, if at least
eight fermions are contracted due to Riemann identities. 

For non-oriented surfaces 
contractions between
chiral and anti-chiral fermions are allowed
$\la \psi(z)\bar \psi(\bar w)\ra_{\sigma}=G_F(z,I_{\sigma}(w))$
with the involution
$I_{\Ac}(w)=I_{\Mc}(w)=I_{\Kc}-{\tau\o 2}=1-\bar w$
and $G_F(z,w)={i\o 2}{\theta_{\alpha}(z-w,\tau)\theta_1'(\tau)\o
\theta_1(z-w,\tau) \theta_{\alpha}(\tau)}$. 
Using \rid\ and taking the sum over
worldsheets $\sigma=\Kc, \Ac, \Mc$  we
finally obtain for the type I one loop   $R^2$ correction 
\eqn\rI{
\Delta^I_{\rm grav}(U)= {V^{8}T_2\o 2^{8} \pi^5}
\Gamma(3) c_0{U_2^3\o T_2^3}\sum_{j,p\neq 0} {1\o |j-p U|^6}\ ,}
where  $c_0={N^2-N+2^4\o 2}$.
The coupling coincides with 
the one--loop correction to $F_1^2 F_2^2$ in eq. \degTTTT. 
In the decompactification limit 
$\Delta_{\rm grav}^I$ disappears,
in agreement with heterotic-type I duality in ten dimensions \ts.
The heterotic tree-level $R^2$ term is 'dual' to the disc diagram on 
the type I side. 
Duality relates this term to a one--loop correction to ${R}^2$ 
on the heterotic side. Since such a term does not exist, we conclude that
on the type I side it is a combination of ${R}^2$ and 
$\Delta_{F_1^2 F_2^2}$,
which corresponds to the heterotic ${R}^2$ term.
In particular this combination is such that no one--loop
correction to ${R}^2$ is predicted on the heterotic side.
A similar observation was recently made with the ${R}^2$ correction
in the duality of heterotic--type IIA \costas.

{\bf Acknowledgements:}
We thank C. Angelantonj, D. Anselmi, I. Antoniadis, C. Bachas, 
W. Lerche, and N.P. Warner for helpful discussions.
St. St. thanks \'Ecole Polytechnique for the warm hospitality.

\goodbreak

\appendix{\appA}{Regularization of world--sheet torus integrals}

The integral for the degenerate orbit of \setup\ and \TODO\ takes the generic 
form:
\eqn\deg{\eqalign{
I_2&=T_2\int_0^\infty{d\tau_2\o \tau_2^{2+\alpha+\epsilon}}\ 
\sum_{(j,p)\neq (0,0)}{d\tau_2\o \tau_2^{2+\alpha+\epsilon}}\ 
e^{-{\pi T_2\o \tau_2U_2}b}\ e^{i\theta (j+\ov Up)}\ c(0)\ \cr
&=c(0){U_2\o\pi}\lf(U_2 \o \pi T_2\ri)^{\alpha+\epsilon}
\Gamma(1+\alpha+\epsilon)\sum_{(j,p)\neq (0,0)}
{e^{i\theta (j+\ov Up)}\o b^{1+\alpha+\epsilon}}\ ,\cr}}
with some:
\eqn\b{\eqalign{
b&=(j+B)^2+C^2\ .\cr}}
We have introduced a regularization $\epsilon$, which is necessary for
$\alpha=0$.
This also allows us to interchange the order of 
summation and integration.
For the sum over $j$ we use the formula ($s\geq 1, C\neq 0$):
\eqn\jsum{\eqalign{
\sum_{j=-\infty}^\infty {e^{i\theta j}\o [(j+B)^2+C^2]^s}&=
{2\pi^s\o\Gamma(s)}{|C|^{s-\h}\o C^{2s-1}}\sum_{r \neq 0}
e^{2\pi i (r-{\theta\o 2\pi}) B}
|r-{\theta\o 2\pi}|^{s-\h}K_{s-\h}\lf[2\pi|C||r-{\theta\o 2\pi}|\ri]\cr
&+\cases{{2^{3/2-s}\pi^{1/2}\o \Gamma(s)}{1\o C^{2s-1}}\ e^{-i\theta B}\  
|\theta C|^{s-1/2}\ K_{s-1/2}(|\theta C|)\ , &$\theta\neq 0$\cr
  {\pi^{1/2}\o\Gamma(s)}\ \Gamma(s-1/2)\ {1\o C^{2s-1}}\ , &$\theta= 0$\ .}
\cr}}
Here $K_\nu(z)$ is the modified Bessel function (of the third kind)
\eqn\bessel{\eqalign{
K_{\nu}(z)&=\h \Gamma(\nu)\Gamma(1-\nu)\lf[I_{-\nu}(z)-I_{\nu}(z)\ri]\ ,\cr
I_\nu(z)&=\sum_{m=0}^\infty{({z\o 2})^{2m+\nu}\o m!\ \Gamma(m+\nu+1)}\ ,\cr}}
while $I_{\nu}(z)$ being the modified Bessel function (of the first kind)
\erdelyi.
For $s=1$ and $0\leq\theta<2\pi,\ C>0$
eq. \jsum\ reduces to the Sommerfeld--Watson transformation.
In fact,  with $K_\h(z)=\sqrt{\pi\o 2x}e^{-x}$, we obtain:
\eqn\sommer{\eqalign{ 
\sum_{j=-\infty}^\infty {e^{i\theta j}\o (j+B)^2+C^2}&=\cr
{\pi\o C} \lf[e^{-i\theta(B-iC)}\ri.&+\lf.e^{-i\theta(B+iC)}{e^{2\pi i(B+iC)}\o 1-e^{2\pi i(B+iC)}}+
e^{-i\theta(B-iC)}{e^{-2\pi i(B-iC)}\o 1-e^{-2\pi i(B-iC)}}\ri]\ .\cr}}
Our formula \jsum\ may be obtained from a generalization of the identity 
(79) given in \van.

In the following we concentrate on the extra contribution to \deg\
originating from the regularization. It arises from the second term in 
\jsum, i.e. it is the contribution from $r={\theta \o 2\pi}$.
We shall denote it by $I_2(\epsilon)$.
The first term in eq. \jsum, given by a sum over integers $r$
and the Bessel function,  has already been evaluated 
in section 2 and leads to the polylogarithms. 

Now we come to the regularization of \setup, which is necessary for 
$2\alpha+N\equiv 2s-2r+N=0$. In that case one has to replace the second sum 
in \Itwoo\ with the expression:
\eqn\finalone{\eqalign{
I_2(\epsilon)={\p^N \o \p\Lambda_1^{q_1} \p\Lambda_2^{q_2}
\p\Lambda_3^{q_3} \p\Lambda_4^{q_4}}&  
{2\o [1-i(\Lambda_1+\Lambda_2+\Lambda_3+\Lambda_4)-{1\o 4}
(\Lambda_1-\Lambda_2+\Lambda_3-\Lambda_4)^2]^{\h(1+2\alpha+2\epsilon)}}\cr
&\times {(2\pi)^\alpha\o \pi^{1/2}}\ \lf(1\o \pi T_2 U_2\ri)^\epsilon\ 
\Gamma(\h+\alpha+\epsilon)\ \zeta(1+2\epsilon)
\lf.\ri|_{\Lambda_i=0}\ .\cr}}

Next, the regularized expression for the $p$--sum in \rd\ is $(k=2g-2)$:
\eqn\finaltwoa{\eqalign{
{2\o(2T_2U_2)^k}&
\sum_{\alpha=0\atop \alpha=\alpha_1+\alpha_2}^{2g}\sum_{s=0}^{k}\ 
(-1)^{s-k}\ \pi^{-\alpha-\h}\ 2^{-4k+2s+\alpha}\ 
{(2k)!\o (2s)! (k-s)!}\lf(1\o 2 T_2U_2\ri)^{\alpha-k}\ 
\lf(1\o \pi T_2U_2\ri)^\epsilon\cr  
&\times\zeta(1-2k+2\alpha+2\epsilon)\ 
\Gamma(\h+s-k+\alpha+\epsilon)\ 
c(0;2g-2\alpha_1,2g-2\alpha_2)\ {(-\pi)^\alpha\o(\alpha_1)!(\alpha_2)!}\ .\cr}}
We have used the relations ($\nu=\h+\alpha+\epsilon$):
\eqn\ident{\eqalign{
{d^{2s}\o dx^{2s}}e^{-ax^2}\lf.\ri|_{x=0}&=
{(2s)! \o (s)!}(-a)^{s}\ ,\cr
{d^{2k}\o dx^{2k}}\ e^{-x}x^\nu K_\nu(x)
\lf.\ri|_{x=0}
&=\sum_{s=0}^k(-1)^{s-k}\ 2^{-2k+2s-\h+\alpha+\epsilon}
\ {(2k)!\o (2s)!(k-s)!}\ \Gamma(\nu+s-k)\ .\cr}}
The sum \finaltwoa\ agrees with eq. (4.24) of \MM\ after performing 
some obvious redefinitions. For $\alpha=k=2g-2$, we encounter 
\eqn\finaltwo{\eqalign{
I_2(\epsilon)&={2\o(2T_2U_2)^k} \sum_{k=\alpha_1+\alpha_2}\sum_{s=0}^{2g-2}\ 
(-1)^{s-k}\ \pi^{-k-\h}\ 2^{-3k+2s}\ {(2k)!\o (2s)! (k-s)!}\cr  
&\times\ \lf(1\o \pi T_2U_2\ri)^\epsilon\zeta(1+2\epsilon)\ 
\Gamma(\h+s+\epsilon)\  
c(0;2g-2\alpha_1,2g-2\alpha_2)\ {(-\pi)^k\o(\alpha_1)!(\alpha_2)!},\cr}}
which must replace the $p$--sum of \rd, when it is  divergent.

To obtain from \finalone\ and \finaltwo\ the $\epsilon$--independent
terms, we have to use the series (cf. also \MM):
\eqn\series{\eqalign{
\zeta(1+2\epsilon)&={1\o 2\epsilon}+\gamma_E+{\cal O}(\epsilon)\ ,\cr
\Gamma(\h+s+\epsilon)&=\Gamma(\h+s)[\psi(\h+s)+\epsilon]+
{\cal O}(\epsilon^2)\ ,\cr
\lf(1\o \pi T_2U_2\ri)^\epsilon&=1-\epsilon\ln(\pi T_2 U_2)+
{\cal O}(\epsilon^2)\ .\cr}}
Here $\gamma_E$ is the Euler--Mascheroni constant $\gamma_E$.

A special case of \finalone\ appears for $N=\alpha=0$, which corresponds to 
the integral:
\eqn\dixon{
\int\limits_{\cal F} {d^2\tau\o \tau_2}\ \sum_{(p_L,p_R)}
\ q^{\h|p_L|^2}\ov q^{\h|p_R|^2}=\kappa-\ln|\eta(T)|^4|\eta(U)|^4\ .}
The integrand has an IR--divergence for $\tau_2\rightarrow\infty$
and thus has to be regularized. This results in an extra contribution, given
by  $\kappa$.
In ref. \DKLII, the integral \dixon\ has been regularized by substracting
the (non--modular invariant) field--theoretical part with the result 
$\kappa=\gamma_E-\ln(4\pi) -\ln(T_2U_2) -1-\ln({2\o 3\sqrt 3})$. 

On the other hand, from our regularization \deg, we obtain:
\eqn\myreg{
I_2(\epsilon)={2\o \pi^\h} 
\lf(1\o \pi T_2U_2\ri)^\epsilon\Gamma(\h+\epsilon)\ \zeta(1+2\epsilon)\ .}
Extracting from $I_2(\epsilon)$ the $\epsilon$--independent terms gives
\eqn\myregg{\kappa=\gamma_E-\ln(4\pi)-\ln(T_2U_2)\ .}

\listrefs

\end